\documentclass[11pt,a4paper]{article}

\usepackage{bm,amsfonts}
\usepackage{amsmath}
\usepackage{graphicx}
\usepackage{hyperref}
\usepackage{color}

\newcommand{\imag}{\mathrm{Im}\,}

\newcommand{\Z}{\mathbb{Z}}

\makeatletter

\@addtoreset{equation}{section}
\makeatother

\date{\today}

\begin{document}

\begin{titlepage}

\renewcommand{\thefootnote}{\fnsymbol{footnote}}

\begin{flushright}
\end{flushright}

\vskip5em

\begin{center}
 {\Large {\bf 
 Hall and spin Hall viscosity in 2d topological systems
 }}

 \vskip2em

 {\sc Taro Kimura}

 \vskip1em

{\it 
Institut de Math\'ematiques de Bourgogne,
Universit\'e Bourgogne Franche-Comt\'e,
France
}


 \vskip3em

\end{center}

 \vskip2em

\begin{abstract}
We study linear responses to metric perturbation in two-dimensional topological systems based on the Dirac fermion formalism.
We introduce a novel quantity, Hall viscosity to particle density ratio, which is analogous to the viscosity to entropy ratio suggested by AdS/CFT correspondence.
This quantity corresponds to the filling fraction which characterizes the quantum Hall states, and hence it could be discussed in Haldane's zero-field quantum Hall system.
We also consider dissipationless viscosity in the time reversal invariant quantum spin Hall system, which is a spin analog of the Hall viscosity.
\end{abstract}

\end{titlepage}


\setcounter{footnote}{0}



\section{Introduction}\label{sec:intro}

Recent interest in topological aspects of electron states has led to a large number of experimental and theoretical works.
In particular, the study of Quantum Hall Effect (QHE)~\cite{PrangeGirvin198912} has shown that the adiabatic geometric phase plays a key role to understand their topological properties.
One of the most remarkable properties in such a topological state is dissipationless transport: 
There is no dissipative longitudinal current, while only a dissipationless transverse current, called the Hall current, is observed in the QH phase.
In a system without time reversal symmetry, we can consider not only such a dissipationless charge transport, but also a dissipationless momentum transport.
The latter is related to a dissipationless viscosity coefficient~\cite{PitaevskiiLifshitz198106,Avron:1995fg}, which is called ``asymmetric viscosity''~\cite{Avron:1995fg}, ``Hall viscosity''~\cite{Read:2008rn,Haldane:2009ke}, and also ``Lorentz shear modulus''~\cite{tokatly:2007lsm,tokatly:2008lsm}.
In order to evaluate the Hall viscosity, we can apply the Kubo formula with correlation functions of the stress tensor in the same way as the dissipative shear viscosity, so that the viscosity indicates a response to a metric perturbation. 
As in the case of the Hall conductivity, the Hall viscosity is also described by the Berry curvature on a certain parameter space, the moduli space of the torus~\cite{Avron:1995fg}, instead of the Brillouin zone for the charge conductivity~\cite{Thouless:1982zz}.
It reflects that the metric of the system can be parametrized by the modulus of the torus.

In earlier works, the Hall viscosity of the lowest Landau level (LLL) has been computed for non-interacting~\cite{Avron:1995fg} and interacting electron states, such as the Laughlin state and generalized conformal blocks~\cite{Read:2008rn,tokatly:2007lsm, tokatly:2008lsm}.
They correspond to the integer and fractional QH states, respectively.
Although the adiabatic property for the higher LL state has been also considered in Ref.~\cite{levay:2792}, the connection with the Hall viscosity has not been explicitly referred.
Furthermore, all the previous results are restricted to the non-relativistic systems although there are various applications of the relativistic Dirac fermion in the context of condensed-matter physics. 
From this point of view, in this paper, we will discuss the Hall viscosity of the two-dimensional Dirac fermion, and explore its applications to graphene~\cite{neto:109} and two-dimensional topological insulators~\cite{Hasan:2010xy}, which have been remarkable topics in recent years.
See also more recent developments in the relation to field theory~\cite{Hughes:2011hv,Nicolis:2011ey,Kimura:2011ef} and gauge/gravity duality~\cite{Saremi:2011ab}.

To discuss topological insulator models, we will start with Haldane's zero-field QH model (Haldane model)~\cite{PhysRevLett.61.2015}.
For the QH system, the filling fraction, which is defined as a ratio of density of states to magnetic flux, is useful to characterize its topological property.
In the topologically non-trivial phase of the Haldane model, however, 
density of states itself vanishes because no net external magnetic field is applied to this model.
Although, in this sense, the filling fraction becomes indeterminate, we can obtain a finite Hall conductivity through an alternative method, called the Widom-St\v{r}eda formula~\cite{Widom1982474,0022-3719-15-22-005}.
We remark that this formula is now extended to the thermal Hall conductivity \cite{Nomura:2011hn} and also the Hall viscosity~\cite{Hidaka:2012rj}.
Similarly, the Hall viscosity of this model also goes to zero even in the topologically non-trivial phase because the viscosity is a hydrodynamic quantity, and should be proportional to density or the magnetic field in this case.
It means that the naive definition of the Hall viscosity cannot capture the topological property of the Haldane model.
From this point of view, we have to consider a novel viscosity, 
which is able to characterize the topological phase of the Haldane model.
We will show a hint to overcome this difficulty can be found in string theory.

Recent progress in string theory has provided us with a remarkable method to explore strongly correlated systems, which is called {\it AdS/CFT correspondence}~(See, e.g., a review article \cite{Aharony:1999ti} for details).
Indeed, there have been a large number of applications discussed in the context of particle physics to condensed-matter physics.
The shear viscosity for a certain strongly coupled system was explored in Ref.~\cite{Policastro:2001yc}, and universal behavior of the viscosity to entropy density ratio, $\eta/s=\hbar/(4\pi k_\mathrm{B})$, was proposed in Ref.~\cite{Kovtun:2004de}.
It suggests that a ratio of viscosity is more fundamental quantity than viscosity itself.
From this point of view, in this paper, we will study a ratio for the Hall viscosity and discuss its properties.

Furthermore, it has been shown that the topological phase can be observed even in time reversal symmetric systems~\cite{PhysRevLett.95.226801}, which is characterized by $\mathbb{Z}_2$ topological invariant~\cite{PhysRevLett.95.146802}.
Its discovery has a significant influence on both of experimental and theoretical sides of condensed-matter physics.
The effective theory of such a system is well described by the Dirac fermion in the vicinity of the avoided crossing point~\cite{1367-2630-12-6-065010}.
Thus, we can also generalize the dissipationless viscosity to the time reversal symmetric system.
As discussed in Ref.~\cite{Avron:1995fg}, the Hall viscosity itself is observed only in the time reversal broken system.
However, we can introduce another kind of viscosity in such a system, which is analogous to the spin Hall conductivity.
We will show that it characterizes a momentum transport in the quantum spin Hall (QSH) system induced by a spin current, while the Hall viscosity is related to the charge current.

The remaining part of this paper is organized as follows.
We begin in Sec.~\ref{sec:visco} with the two-dimensional Dirac fermion under the metric perturbation, which will play a fundamental role in the study of two-dimensional topological insulators.
We then discuss its application to graphene system.
In Sec.~\ref{sec:ratio}, we explore two-dimensional topological insulator models with/without time reversal symmetry.
We introduce a novel quantity, the Hall viscosity ratio, to discuss the topological phases in these models, and analogous properties with a similar quantity, which is proposed in the context of AdS/CFT correspondence.
We also discuss a spin Hall analogue of the viscosity ratio in the time reversal symmetric system.
Sec.~\ref{sec:summary} is devoted to summary and discussions.

\section{Hall viscosity and Dirac fermion}\label{sec:visco}


Let us start with how to calculate the Hall viscosity.
As in the case of the ordinary dissipative shear viscosity, one can apply the linear response theory to the Hall viscosity with the metric perturbation, which is related to the shear stress mode.
We describe the metric perturbation as follows:
\begin{equation}
 g_{ij} = \frac{J}{\tau_2}  \left(
 \begin{array}{cc}
  1 & \tau_1 \\
  \tau_1 & \left|\tau\right|^2 \\
 \end{array}
 \right), \qquad
 g^{ij} = \frac{1}{J\tau_2} \left(
 \begin{array}{cc}
  \left|\tau\right|^2 & -\tau_1 \\
  -\tau_1 & 1 \\
 \end{array}
 \right),
 \label{metric01}
\end{equation}
where $g^{ij}$ stands for the inverse of $g_{ij}$, and $\tau=\tau_1+i\tau_2 \in \mathbb{C}$ is a modulus of the torus, which plays a role of the perturbation parameter.
We obtain the Euclidean metric by setting $\tau_1=0$, $\tau_2=1$ ($\tau = i$) and $J=1$.
Then, the stress tensor is given as a response to the metric perturbation, 
$T_{ij} = -2 \left\langle \partial H/\partial g_{ij} \right\rangle$.
According to the linear response theory, one obtains the shear viscosity from a correlation function of the stress tensor, $\left\langle \left[T_{xy},T_{xy}\right] \right\rangle$.
In addition to the dissipative shear and bulk viscosity, one can consider another transport coefficient in the time reversal broken system~\cite{PitaevskiiLifshitz198106}, which is called the {\it Hall viscosity}.

The Hall viscosity is also given by a correlator of stress tensors, such as
$\left\langle \left[T_{xx},T_{xy}\right] \right\rangle$.
In fact, it is shown in Ref.~\cite{Avron:1995fg} that it is also related to the Berry curvature on the moduli space of the metric perturbation parameters $(\tau_1,\tau_2)$,
\begin{equation}
 \eta_\mathrm{H}^{(\alpha)}
 = 8\ \imag \left\langle \, \frac{\partial \Psi_\alpha}{\partial g^{xx}}
 \, \Bigg| \, \frac{\partial \Psi_\alpha}{\partial g^{xy}} \, \right\rangle
 = 2\ \imag \left\langle \, \frac{\partial \Psi_\alpha}{\partial \tau_1}
 \, \Bigg| \, \frac{\partial \Psi_\alpha}{\partial \tau_2} \, \right\rangle.
\end{equation}
This is a contribution of the eigenstate $\left|\Psi_\alpha\right\rangle$ to the viscosity, measured with the unit $\hbar/L^2$ with the system size $L$, and $\alpha$ denotes a label of the particle state, i.e., the Landau level index.
In order to obtain the total contribution, we should take into account all the eigenstates under the Fermi energy.
We emphasize that the Hall viscosity is associated with the curvature on the specific point in the moduli space.
Therefore, it is not quantized although the integral of the curvature over the moduli space is quantized~\cite{Avron:1995fg}.
This is in contrast to the Hall conductivity, which is given by integral of the Berry curvature over the Brillouin zone, and therefore quantized~\cite{Thouless:1982zz}.

\subsection{Dirac fermion}\label{sec:Dirac}

In order to discuss the topological insulator models, we study the Hall viscosity of the two-dimensional Dirac fermion model with external magnetic field.

We start with the two-dimensional Dirac Hamiltonian in the form of
\begin{equation}
 \mathcal{H} = v \left( p_x \sigma_x + p_y \sigma_y \right) + m \sigma_z,
 \label{Dirac01}
\end{equation}
where $v$ is the Fermi velocity, $p_{x,y}$ is the momentum, $m$ is the mass, and $\sigma_{x,y,z}$ are the Pauli matrices.
We focus on the massless case $m = 0$ for the moment.
Let us examine a behavior of the Dirac Hamiltonian (\ref{Dirac01}) under the metric perturbation (\ref{metric01}). 
For this purpose, we rewrite the metric (\ref{metric01}) in the ``zweibein'' formalism, $g_{ij} =
e_i^a e_j^b \delta_{ab}$, where $\delta_{ab}$ is the Euclidean metric, as follows:
\begin{equation}
 g_{ij} = \frac{J}{\tau_2}
 \left(
 \begin{array}{cc}
  1 & 0 \\
  \tau_1 & \tau_2 \\
 \end{array}
 \right) \left(
 \begin{array}{cc}
  1 & 0 \\
  0 & 1 \\
 \end{array}
 \right) \left(
 \begin{array}{cc}
  1 & \tau_1 \\
  0 & \tau_2 \\
 \end{array}
 \right).
\end{equation}
The modified Pauli matrices, satisfying $\left\{\tilde \sigma_i, \tilde\sigma_j \right\}=2g_{ij}$, are obtained through the linear transformation via the zweibein,
\begin{equation}
 \left(
 \begin{array}{c}
  \tilde \sigma_x \\ \tilde\sigma_y
 \end{array}
 \right) = \sqrt{\frac{J}{\tau_2}} \left(
 \begin{array}{cc}
  1 & 0 \\
  \tau_1 & \tau_2 \\
 \end{array}
 \right) \left(
 \begin{array}{c}
  \sigma_x \\ \sigma_y
 \end{array}
 \right).
\end{equation}
Hence, the Dirac Hamiltonian with the metric (\ref{metric01}) yields
\begin{equation}
 \mathcal{H}
 = v g^{ij} p_i \tilde \sigma_j 
 = \frac{v}{\sqrt{J\tau_2}} \left( 
 \begin{array}{cc}
  0 & i\bar\tau p_x - i p_y \\
  -i\tau p_x + i p_y & 0 \\
 \end{array}
 \right).
 \label{Dirac02}
\end{equation}
This is the Dirac Hamiltonian in a flat, but with a non-standard metric.
See, e.g.~Ref.~\cite{ParkerToms200908}.

To discuss the Landau quantization of this model, let us take into account the magnetic field.
We introduce a modified momentum $p_j=-i\partial_j+A_j$ with the Landau gauge $A=(-By,0)$.
Then, the Hamiltonian is given in the form of
\begin{equation}
 \mathcal{H} = \omega \left(
 \begin{array}{cc}
  0 & a \\
  a^\dag & 0 \\
 \end{array}
 \right),
 \label{Dirac03}
\end{equation}
where the cyclotron frequency is given by $\omega = v \sqrt{2|eB|/J}$, and the creation and annihilation operators are defined as
\begin{equation}
 a = \frac{1}{\sqrt{2\tau_2|B|}} \left( i \bar\tau p_x - i p_y \right),
  \qquad
 a^\dag = \frac{1}{\sqrt{2\tau_2|B|}} \left( - i \tau p_x + i p_y \right).
 \label{ladder01}
\end{equation}
We see that they satisfy the ordinary bosonic commutation relation, $\left[a,a^\dag\right]=1$.

We explore the eigenstates of the Hamiltonian (\ref{Dirac02}).
The Dirac zero mode, corresponding to the LLL state, is described by the equation, $a \phi_\alpha^{(0)}=0$.
Here $\alpha$ labels the degenerated LLL states, $\alpha=1,\cdots, N$, where the magnetic field is quantized as $\displaystyle 
\frac{|eB|}{2\pi \hbar} L^2 = N$.
The higher LL states are generated by the creation operator,
\begin{equation}
 \phi^{(n)}_\alpha = \frac{1}{\sqrt{n!}} \left(a^\dagger\right)^n
  \phi^{(0)}_\alpha.
\end{equation}
Hence, the eigenvalues and the eigenstates of the Dirac Hamiltonian yield
\begin{subequations}
\begin{align}
 \epsilon^{(\pm n)} & = \pm \omega \sqrt{n} 
 \qquad \mbox{for} \quad n = 0, 1, 2, \ldots,
 \\
 \Phi^{(0)}_\alpha & = \left(
 \begin{array}{c}
  0 \\ \phi^{(0)}_\alpha
 \end{array}
 \right) 
 \qquad \mbox{for} \quad n = 0, 
 \label{eigen01}
 \\
  \Phi^{(\pm n)}_\alpha & =
 \frac{1}{\sqrt{2}} \left( 
 \begin{array}{c}
  \pm \phi^{(n-1)}_\alpha \\
  \phi^{(n)}_\alpha
 \end{array}
 \right) 
 \qquad \mbox{for} \quad n = 1,2,\ldots,
 \label{eigen02}
\end{align}
\end{subequations}
where the normalization condition of the LL state is given by $\left\langle\phi_\alpha^{(n)} \, \Big|\, \phi_\beta^{(n)}\right\rangle= \delta_{\alpha\beta}$.
In the presence of the mass term as in \eqref{Dirac01}, the LL spectrum is slightly modified~\cite{Jackiw:1984ji}
\begin{align}
    \epsilon^{(n)} = 
    \begin{cases}
    \pm \sqrt{\omega^2 n + m^2} & (n = \pm 1, \pm 2, \ldots) \\
    \pm m \operatorname{sgn}(eB) & (n = 0)
    \end{cases}
    \label{LL_massive}
\end{align}
where the sign of the zero-mode energy $\epsilon^{(n = 0)}$ depends on the chirality.
As discussed in Ref.~\cite{levay:2792}, the LLL wavefunction can be written in a form of $\phi_\alpha^{(0)}=\mathcal{N}_\alpha^{-1/2}\psi_\alpha$ where $\psi_\alpha$ is given by the theta function,
\begin{subequations}
\begin{align}
 \psi_\alpha(z) & = 
  e^{-(\pi N/ \imag \tau)(\imag z)^2} 
  \vartheta \left[ \begin{array}{c} \alpha/N \\ 0 \end{array} \right]
  (Nz|N\tau),
    \\
 \vartheta \left[ \begin{array}{c} \alpha/N \\ 0 \end{array} \right]
  (Nz|N\tau)
   & = \sum_{m\in\Z}
   e^{i\pi(m+\alpha/N)^2 N\tau + 2i\pi(m+\alpha/N)Nz} ,
\end{align}
\end{subequations}
with letting $z = x + i y$ be the complex coordinate in two dimensions.

The property of the theta function shows the normalization constant
$\mathcal{N}_\alpha=(\tau_2)^{-1/2}$.
It depends on both of $\tau$ and $\bar\tau$ because it is also written
as $\tau_2 = \imag \tau = (\tau - \bar\tau)/2i$.
In general, the Berry curvature on the moduli space is related to the
holomorphic structure of the eigenstate.
Actually the LLL state splits into the (anti-)holomorphic and
non-holomorphic parts, corresponding to the theta function and the
normalization factor.
It is then shown that the contribution to the curvature comes
only from the normalization
factor~\cite{Read:2008rn,tokatly:2007lsm,tokatly:2008lsm}.
According to the Berry curvature of the LLs given in
Ref.~\cite{levay:2792}, we obtain the curvature of the eigenstate of
(\ref{Dirac03}),
\begin{subequations}\label{curvature03}
\begin{align}
 \mathcal{F}^{(0)}_{\alpha\beta} & = - \frac{1}{4} \frac{d\tau_1
 \wedge d\tau_2}{\tau_2^2} \delta_{\alpha\beta}
 \hspace{1.6em} \mbox{for} \quad n = 0, 
 \label{curvature01}
 \\
 \mathcal{F}^{(n)}_{\alpha\beta} & = - \frac{\left|n\right|}{2} \frac{d\tau_1
 \wedge d\tau_2}{\tau_2^2} \delta_{\alpha\beta}
 \quad \mbox{for} \quad n = \pm 1,\pm 2,\cdots,
 \label{curvature02}
\end{align}
\end{subequations}
where the curvature is the two-form defined as $\mathcal{F}^{(n)}_{\alpha\beta}=d\mathcal{A}_{\alpha\beta}^{(n)}$ with the Berry connection $\mathcal{A}_{\alpha\beta}^{(n)}=i(\Phi_\alpha^{(n)})^\dag d \Phi_\beta^{(n)}$.
Therefore, the Hall viscosity is obtained from this curvature on the moduli space by taking into account the degeneracy of each LL state.
The LL states are $N$-fold degenerated, except for the zero mode which turns out to be half filled because of the parity anomaly~\cite{PhysRevLett.53.2449}.
Thus, contributions to the Hall viscosity obtained from the curvature (\ref{curvature03}) are given by
\begin{equation}
 \eta_\mathrm{H}^{(n)} = \left\{ 
 \begin{array}{cl}
  \hbar \rho /8 & (n=0) \\
  \left|n\right| \hbar \rho /2 & (n=\pm 1,\pm 2,\cdots) \\
 \end{array}
 \right.
 \label{visco01}
\end{equation}
where $\rho=N/L^2=1/(2\pi \ell_B^2)$ is the density of particles, and the magnetic length is defined as $\ell_B=\sqrt{\hbar/|eB|}$.
We remark that the result (\ref{visco01}) depends on the level index $n$.
This is not the case for the Hall conductivity.
One can show that this level dependence indeed reproduces the classical result $\eta_\mathrm{H} = k_{\mathrm{B}}T\rho/(2\omega_c)$, where $\omega_c$ is the cyclotron frequency~\cite{PitaevskiiLifshitz198106,Read:2008rn}.

\subsection{Application to graphene}\label{sec:graphene}

One of the most desirable applications of the relativistic $(2+1)$-dimensional Dirac fermion theory in condensed-matter systems would be the graphene system~\cite{PhysRevLett.53.2449,neto:109}.
The low energy behavior of graphene is well described by two Dirac fermions at two valleys, $K$ and $K'$ points, and each valley has {\it up} and {\it down} spin degrees of freedom.
Hence, we often treat it as the four flavor, totally eight component, Dirac fermion system.
Indeed the anomalous QHE of graphene is well explained by the four flavor theory~\cite{Gusynin:2005pk}.

We shall discuss the Hall viscosity of the graphene system.
First of all, it should be noted, as discussed in Ref.~\cite{Read:2008rn}, that it would be ambiguous for the system where translational symmetry is violated at short length scales, since the viscosity is the transport coefficient related to the momentum transport.
Therefore, we have to consider the system which is well described by the continuum effective theory, i.e., the magnetic length, which characterizes an extent of the wavefunction, is much larger than the lattice spacing $a_L$ of the system.
Furthermore, in the case of graphene, we should also deal with the low energy region where the Dirac fermion description is suitable.


We introduce the effective Hamiltonian for the monolayer
graphene~\cite{PhysRevLett.53.2449},
\begin{equation}
 \mathcal{H}_\mathrm{m} = \omega_\mathrm{m} \left( 
 \begin{array}{cc}
  0 & \pi \\
  \pi^\dagger & 0 \\
 \end{array}
 \right)
 \label{monolayer01}
\end{equation}
where the cyclotron frequency of this system is defined as
$\omega_\mathrm{m} = \sqrt{2}(\sqrt{3}/2)a_L \gamma_{AB}/(\hbar\ell_B)$,
and $\gamma_{AB}$ is the hopping amplitude between the nearest neighbor sites.
Here the annihilation/creation operators are defined as
$\pi=(i\xi\bar\tau p_x-ip_y)/\sqrt{2\tau_2 |eB|}$, $\pi^\dagger=(-i\xi\tau p_x+ip_y)/\sqrt{2\tau_2 |eB|}$ satisfying $\left[\pi,\pi^\dagger\right]=\xi$ where $\xi=1$ and $\xi=-1$ refer to the $K$ and $K'$ valleys, respectively.
This agrees with the model discussed in Sec.~\ref{sec:Dirac}, so that one can apply the Dirac fermion analysis to the current situation with the metric perturbation.
Considering valley and spin degrees of freedom, the Hall viscosity of the graphene system yields $ \eta_\mathrm{H}= \pm 4 \left( |n|(|n|+1)/4 + 1/8 \right) \hbar \rho$ for $n=0, \pm 1,\pm 2,\ldots$. 
The sign corresponds to whether it is associated with particle or hole states, similarly to the anomalous Hall conductivity of the graphene~\cite{Gusynin:2005pk}.
On the other hand, in the non-interacting and non-relativistic system, it becomes $\eta_\mathrm{H}=n^2\hbar\rho/4$ for $n=1,2,\ldots$, since the contribution of each LL is given by $\eta_\mathrm{H}^{(n)}=(n+1/2)\hbar\rho/2$ for $n=0,1,\ldots$, as shown in Refs.~\cite{Read:2008rn,levay:2792}.

\section{Hall and spin Hall viscosity ratio}\label{sec:ratio}

In this Section, we apply the Hall viscosity analysis to two-dimensional topological insulators~\cite{Hasan:2010xy}.
All the topological insulator models have a band gap, which is generated
by magnetic field in a time reversal symmetry broken system, and
spin-orbit interaction in a time reversal symmetric system, respectively.
Although the origins of these gaps are different, topological aspects
of such systems are well described by the effective Dirac fermion model
discussed in Sec.~\ref{sec:visco}.

\subsection{Hall viscosity ratio in topological phase}

First, we consider the model proposed by Haldane~\cite{PhysRevLett.61.2015}, which is a tight-binding model defined on a honeycomb lattice with a complex flux.
The remarkable property of this model is that the topological phase can be observed due to the violation of the time reversal symmetry even though no net external magnetic field is applied.
Indeed it is shown that anomalous contribution of the Dirac zero mode, relating to the parity anomaly, plays an important role in this model.

Let us discuss how to obtain the Hall conductivity for the model without external magnetic field.
It is given by a derivative of charge density with respect to magnetic flux, $\sigma_\mathrm{H}=\partial\rho_c/\partial B$, which is called the Widom-St\v{r}eda formula~\cite{Widom1982474,0022-3719-15-22-005}, which is available even in the zero field limit.
It can be interpreted as l'Hopital's rule to evaluate an indeterminate form of the filling fraction in the limit $B\to 0$.
We calculate the Hall conductivity with the Widom-St\v{r}eda formula after taking into account infinitesimal magnetic field as a probe.
Then, taking the limit $B\to 0$, we obtain the Hall conductivity with no external magnetic field.

We apply this argument to the Haldane model~\cite{PhysRevLett.61.2015}, 
\begin{align}
    \mathcal{H}_\text{H} = \sum_{j=1}^3 d_j(p) \sigma_j
\end{align}
with the coefficients given by
\begin{subequations}
\begin{align}
    d_1(p) & = 1 + \cos p_1 + \cos p_2
    \, , \\
    d_2(p) & = \sin p_1 + \sin p_2
    \, , \\
    d_3(p) & = m + 2 t \left[ \sin p_1 - \sin p_2 - \sin (p_1 - p_2) \right]
    \, .
\end{align}
\end{subequations}
where $m$ and $t$ are the mass and the hopping parameters.
We may write the momenta $p_{1,2}$ in terms of of the orthogonal basis, $\displaystyle p_1 = \frac{1}{2} p_x + \frac{\sqrt{3}}{2} p_y$, $\displaystyle p_2 = - \frac{1}{2} p_x + \frac{\sqrt{3}}{2} p_y$.
Expanding the Hamiltonian around the two valley points, $\displaystyle K: (p_1,p_2) = \left( + \frac{2\pi}{3} + k_1, - \frac{2\pi}{3} + k_2 \right)$ and $\displaystyle K': (p_1,p_2) = \left( - \frac{2\pi}{3} + k_1 , + \frac{2\pi}{3} + k_2 \right)$, we obtain the effective Dirac Hamiltonians,
\begin{subequations}
\begin{align}
    \mathcal{H}_K & = - \frac{\sqrt{3}}{2} (k_x \sigma_x + k_y \sigma_y) + \tilde{m}_- \sigma_z
    \, , \\
    \mathcal{H}_{K'} & = + \frac{\sqrt{3}}{2} (k_x \sigma_x - k_y \sigma_y) + \tilde{m}_+ \sigma_z
    \, ,
\end{align}
\end{subequations}
where the valley dependent mass terms are given by $\tilde{m}_\pm = m \mp 3 \sqrt{3} t$.
We then turn on the probe magnetic field $B$.
Since they have different chiralities for $K$ and $K'$ points, the LL spectrum~\eqref{LL_massive} is given by
\begin{align}
    K: \
    \epsilon^{(n)} = 
    \begin{cases}
    \pm \sqrt{\omega^2 n + \tilde{m}_-^2} \\ 
    - \tilde{m}_- \operatorname{sgn}(eB) 
    \end{cases}
    \quad
    K': \
    \epsilon^{(n)} = 
    \begin{cases}
    \pm \sqrt{\omega^2 n + \tilde{m}_+^2} & (n = \pm 1, \pm 2, \ldots) \\
    + \tilde{m}_+ \operatorname{sgn}(eB) & (n = 0)
    \end{cases}
\end{align}
In the time reversal symmetric case $t = 0$, the masses $\tilde{m}_\pm$ have the same sign, and the contributions of $K$ and $K'$ points are symmetric under $B \leftrightarrow -B$.
Therefore, there is no Hall conductivity $\sigma_\text{H} = 0$ in this case.
If $\tilde{m}_\pm$ have the opposite sign, it is not symmetric under $B \leftrightarrow - B$ for $K$ and $K'$ points, so that there is an extra charge density induced by the magnetic field, $\rho_c = \pm e^2 B / h$, which leads to the Hall conductivity $\sigma_\text{H} = \pm e^2 / h$.
To summarize, the Hall conductivity is given by $\sigma_\text{H} = \nu e^2 / h$ in the zero field limit $B \to 0$ with the index $\displaystyle \nu = \frac{1}{2}\left( \operatorname{sgn}(\tilde{m}_-) - \operatorname{sgn}(\tilde{m}_+) \right)$, which implies that the zero mode is half filled.


We turn to the Hall viscosity in the Haldane model.
As in the case of the Hall conductivity, violation of time reversal symmetry is crucial to obtain the Hall viscosity.
Since only the zero mode contributes to the violation of time reversal symmetry in the zero magnetic field limit $B \to 0$, the Hall viscosity of the Haldane model is given by $\eta_\text{H} = \nu (1/4) \hbar \rho$, where the contribution of each zero mode is given by $\pm (1/8) \hbar \rho$ as shown in Sec.~\ref{sec:visco}.
We remark that not charge, but particle number density is used in the expression of the Hall viscosity (\ref{visco01}).
However, taking the zero field limit, the Hall viscosity goes to zero as the particle density vanishes.
This reflects that viscosity is the hydrodynamical quantity which should depend on the density.
In the Haldane model, the incompressible fluid state of the LLs cannot be observed, and thus the naive definition of the Hall viscosity cannot capture the topological property of the model.

In order to overcome the difficulty in considering viscosity for a system without the LLs, we introduce a novel quantity, a ratio of the Hall viscosity to particle density,
\begin{equation}
 \hat\eta_\mathrm{H}=\frac{\eta_\mathrm{H}}{\rho}.
 \label{visco02}
\end{equation}
This is similar to {\it kinematic viscosity}, which is defined as the ratio of viscosity to mass density.
In fact, the topological phase has an energy gap, which is interpreted as a mass parameter.
In this sense, it seems sufficient to apply the kinematic viscosity to characterize the topological phase.
However, since the unit of viscosity is $\hbar$ times density of particles, the ratio of viscosity to density simply has the dimension of $\hbar$.
From this point of view, the viscosity ratio (\ref{visco02}) should be independent of the characteristic scale of the system, and it seems a universal quantity rather than the kinematic viscosity, which may have a system dependent non-universal mass scale.

With emphasis on its similarity to the Widom-St\v{r}eda formula, we introduce another expression,
\begin{equation}
 \hat\eta_{\mathrm{H}} = \frac{\partial \eta_\mathrm{H}}{\partial \rho} .
 \label{eq:WSformula}
\end{equation}
This is regarded as an infinitesimal analog of the expression (\ref{visco02}).
At least, in the linear response regime, these two expressions are consistent.
By using l'Hopital's rule as in the case of the Hall conductivity, we see that it takes a non-zero value,
\begin{equation}
 \hat\eta_\mathrm{H} = \pm \frac{\hbar}{4}.
 \label{visco03}
\end{equation}
This expression is available even in the zero field limit since there is no $B$ dependence.
Therefore, it can be interpreted as an alternative formula for the Hall viscosity, which is associated with the Widom-St\v{r}eda formula.
In general, the ratio is given by $\hat\eta_\mathrm{H} = s \hbar/2$ with effective one-particle spin $s$ because the Hall viscosity can be represented as $\eta_\mathrm{H}=s\hbar\rho/2$ \cite{Read:2008rn}.

Let us comment on the universality of the Hall viscosity for the LLL state.
The Hall viscosity of the LLL state, discussed in Refs.~\cite{Avron:1995fg,tokatly:2007lsm}, and of the Dirac zero mode shown in (\ref{visco01}) are given by $\eta_\mathrm{H}=\hbar\rho/4$, yielding $\hat\eta_\mathrm{H}=\hbar/4$.
Furthermore, it is shown in Refs.~\cite{Read:2008rn,tokatly:2008lsm} that the
Hall viscosity of $\nu=1/m$ Laughlin state becomes $\eta_\mathrm{H} = m \hbar
\rho/4$ with $\rho=1/(2m\pi\ell_B^2)$.
Although the value of the Hall viscosity itself is different from the
above, dependence on the magnetic field is universal,
$\eta_\mathrm{H}/B=e/8\pi$.
On the other hand, this universality is not observed in the excited state.
Indeed, the Hall viscosity for more generalized conformal blocks is obtained in Ref.~\cite{Read:2008rn}, $\eta_\mathrm{H}=(\nu^{-1}/2+h_\psi)\hbar\rho/2$, with a conformal weight $h_\psi$.
Similarly, the $n$-th LL state gives rise to $\eta_\mathrm{H}^{(n)} = (n+1/2)\hbar\rho/2$ as shown in Refs.~\cite{Read:2008rn,levay:2792}.
These results yield a larger value $\eta_\mathrm{H}/B>e/8\pi$, compared with the LLL state, since an excited state has a higher spin.

We also remark similarities between the Hall viscosity ratio $\hat\eta_\mathrm{H}$ and another quantity, the viscosity to entropy density ratio, $\eta/s$, which is well studied in the context of AdS/CFT correspondence:
The dissipative shear viscosity in a strongly coupled system is obtained in Ref.~\cite{Policastro:2001yc} based on AdS/CFT correspondence, and its universal behavior, $\eta/s=\hbar/(4\pi k_\mathrm{B})$, called {\it KSS bound}, is then discussed in Ref.~\cite{Kovtun:2004de}.
In the context of AdS/CFT correspondence, the Hawking temperature of the black hole is interpreted as the temperature of the dual field theory at boundary.
Although the temperature goes to zero in the extremal black hole limit, the residual entropy is observed in several cases.
Actually it is discussed in Refs.~\cite{Chakrabarti:2009ht,Paulos:2009yk} that this universality is still available even in such an extremal case, corresponding to the zero temperature limit.
This extremal limit is analogous to the zero magnetic field limit in our model because the ratio can be indeterminate in the limit.
It suggests that such a quantity could be available even for the extremal case.

It is shown in Ref.~\cite{Brigante:2007nu} that the KSS
bound can be violated in generalized holographic model,
e.g., Gauss-Bonnet gravity, which includes higher order curvature terms.
In the case of the Hall viscosity, the shift of its value due to the geometry effect has been pointed out~\cite{Read:2008rn}.
These results suggest that we should carefully incorporate the geometric effect in the study of the viscosity.

We remark that these universal behaviors are not discussed for viscosity itself, but for the viscosity ratios.
The reason why is the natural unit of viscosity should be $\hbar$ times density as discussed in Ref.~\cite{Read:2008rn}.
This density dependence suggests viscosity relates to the characteristic scale of the system, e.g., the magnetic length.
From this point of view, we should consider the ratio of quantities rather than viscosity itself to discuss the universality of the viscosity.

\subsection{Spin Hall viscosity with time reversal symmetry}

It is well known that the Hall conductivity is observed if and only if the time reversal symmetry is broken.
As in the case of conductivity, it is pointed out in Ref.~\cite{Avron:1995fg} that the Hall viscosity can be also observed in a time reversal symmetry broken system.
On the other hand, the spin Hall conductivity induced by a spin-orbit interaction is found in time reversal symmetric
systems~\cite{murakami-2003-301,PhysRevLett.92.126603} with no external magnetic field applied.
In this subsection, we consider a spin Hall analogue of viscosity, and study its behavior in the QSH system.

The most fundamental model for the two-dimensional topological insulator with time reversal symmetry (QSH system) is the Kane-Mele model, which has been proposed in Refs.~\cite{PhysRevLett.95.226801,PhysRevLett.95.146802}.
In this model, the spin-orbit interaction is taken into account on the honeycomb lattice instead of magnetic field, which corresponds to the next nearest neighbor (NNN) hopping term of the Haldane model.
Actually either up or down spin sector of the Kane-Mele model is equivalent to the Haldane model.
However, although the time reversal symmetry of the Haldane model is explicitly broken, the Kane-Mele model preserves the time reversal symmetry by combining both up and down spin sectors.
In this sense, the Kane-Mele model can be interpreted as the {\it doubled} Haldane model where the signs of the NNN hopping amplitude is opposite for up and down spins.
If we do not incorporate any other spin-mixing terms, a spin current is derived from the quantized Hall current of the Haldane model, $J_s=(\hbar/2e)(J_\uparrow - J_\downarrow)$, yielding a spin Hall conductivity $\sigma_{\mathrm{sH}}=e/2\pi$.
From this point of view, we introduce a spin Hall analogue of the Hall viscosity, that we call {\it spin Hall viscosity},
\begin{align}
    \eta_\text{sH} = \frac{1}{2} \left(\eta_\mathrm{H}^{(\uparrow)} - \eta_\mathrm{H}^{(\downarrow)} \right)
    \, .
\end{align}
The spin Hall viscosity shall be proportional to the particle density, so that it would be zero in the Kane-Mele model as in the case of the Haldane model.
Therefore, we define an alternative quantity, {\it spin Hall viscosity ratio}, in the topological phase of the Kane-Mele model,
\begin{equation}
 \hat{\eta}_\text{sH} = \frac{\partial \eta_\text{sH}}{\partial \rho} 
 = \frac{1}{4}\hbar.
 \label{visco04}
\end{equation}
As in the case of the Haldane model, the quantity defined in
(\ref{visco04}) is for the one-particle state since it is divided by the particle density.
We remark that it is not derived from hydrodynamical relations, however such a spin-related viscosity would be well-defined unless spin-mixing terms, which do not preserve $z$-component of spin, are not imposed.
In this sense, as well as the spin Hall conductivity, this quantity would not be directly related to a generic topological invariant, which characterizes the $\mathbb{Z}_2$ topological phase.
Let us comment many-body interaction effect on this quantity.
Although some generalizations of topological insulators for correlated electron systems have been considered~\cite{PhysRevLett.102.256403,pesin-2009}, an explicit consequence is not yet obvious.
If certain aspects of highly correlated topological insulators are understood, we would be able to discuss the spin Hall viscosity for such systems.

\begin{figure}[t]
 \begin{center}
  \includegraphics[width=11.6em]{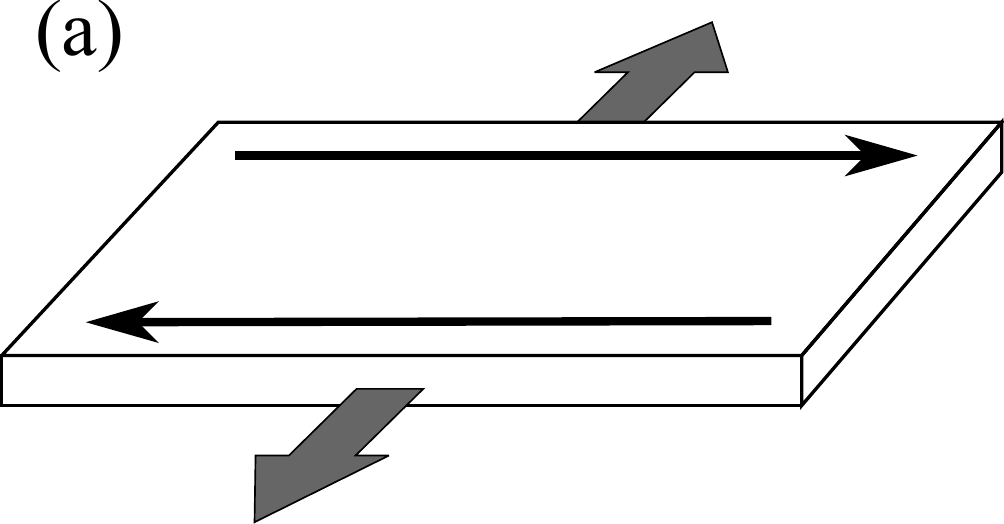} \qquad
  \includegraphics[width=11.6em]{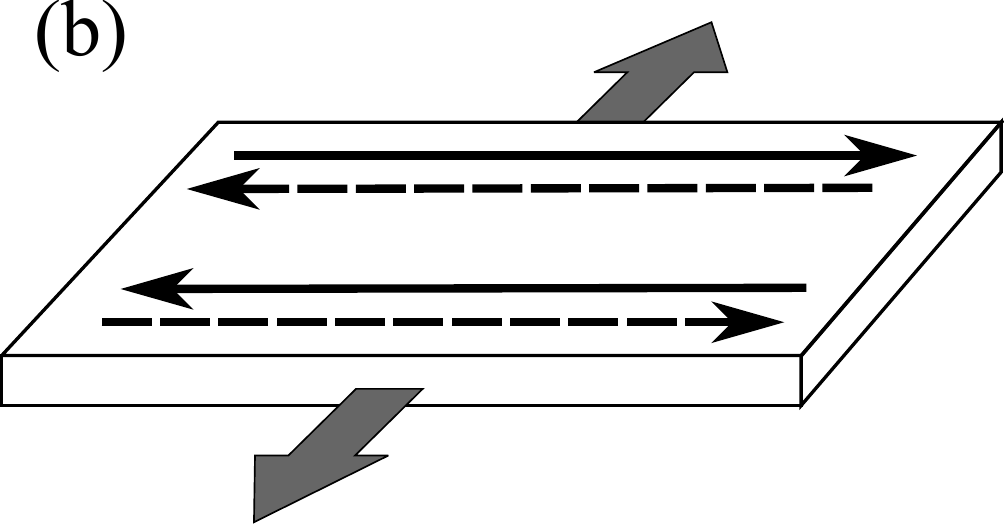}
 \end{center}
 \caption{Stress per particle $\hat{T}_{ij}$ induced by (a) chiral current in QH state and (b) helical
 current (spin current) in QSH state. 
 Solid (dotted) lines represent currents of up (down) spin.}
 \label{stress_fig}
\end{figure}

Let us discuss a physical meaning of the spin Hall viscosity.
In the presence of magnetic field, the Hall viscosity gives a relation between the electric current and the stress tensor, e.g., $T_{xx}=-T_{yy}=\eta_\mathrm{H}\partial_y v_x$, where $v_x$ stands for the current in $x$-direction~\cite{tokatly:2008lsm}.
In the case of the spin Hall system, it corresponds to $T_{xx}=-T_{yy}=\eta_{\mathrm{sH}}(\partial_y v_{x}^{(\uparrow)} - \partial_y v_{x}^{(\downarrow)})=\eta_{\mathrm{sH}}\partial_y v_{s,x}$ where $v_{s,x}$ represents the $x$-component of a spin current.
We remark that a naive stress tensor itself would be zero in the Kane-Mele model since $\eta_\text{sH} \to 0$ as discussed earlier.
Therefore, we should instead consider the stress per particle, $\hat{T}_{ij} = \partial T_{ij}/\partial \rho$, in this case.
See \cite{Hidaka:2012rj} for a related discussion.
Then, the spin current induces the stress in the Kane-Mele model as shown in Fig.~\ref{stress_fig}(b), similarly to the chiral current in the QH state (Fig.~\ref{stress_fig}(a)).
It is dissipationless because the stress is perpendicular to the current as in the cases of Hall and spin Hall conductivity.
It would be possible to detect such dissipationless viscosities, for example, using the piezoelectric effect, which converts the stress to the electric signal.


\section{Discussions}\label{sec:summary}

In this paper, we have discussed the Hall and spin Hall viscosity in two-dimensional topological insulator models.
We have studied the Dirac fermion model under the metric perturbation.
The Berry curvature on the moduli space, which gives rise to the Hall viscosity of the Dirac fermion, is computed, and its application to the graphene system is also addressed.
We have then proposed the novel quantity, the Hall viscosity to particle density ratio, to study the viscosity of Haldane's zero-field model for QHE.
It is analogous to the ratio of viscosity to entropy density, which is often explored in the context of AdS/CFT correspondence, and similar properties are discussed.
We have also studied the spin Hall viscosity of the Kane-Mele model.
Although the Hall viscosity is observed only in the time reversal broken systems, the spin Hall viscosity can be found even in the time reversal symmetric systems.
We have pointed out that the spin current induces the dissipationless stress in the topological phase of the Kane-Mele model.

We now comment some open problems along this direction.
One of them is the relation between the Hall viscosity and the edge state.
It is well known that the edge state plays an important roll on the transport phenomena both in the QH and QSH systems: 
The Hall conductivity is actually obtained through the bulk/edge correspondence~\cite{PhysRevLett.71.3697}.
In this sense, the Hall viscosity is expected be discussed from the edge state point of view as well.
Next is a generalization to the three dimensional topological insulators, which are also characterized by $\mathbb{Z}_2$ invariant~\cite{fu:106803,PhysRevB.75.121306,PhysRevB.79.195322}.
Three-dimensional anomalous viscosity induced by the external field was indeed considered in Ref.~\cite{PitaevskiiLifshitz198106}, and its classical computation was also given.
Since the surface state of the three-dimensional model can be described by Dirac fermions, the results discussed in this paper would be similarly applied.
Furthermore, the four dimensional generalization would be also an interesting direction.
The anomalous contribution in the four dimensional hydrodynamics is recently considered in Ref.~\cite{Son:2009tf}.
It is related to the triangle anomaly, and corresponds to the non-linear response coefficient.
In the four dimensional QH system~\cite{Zhang:2001xs}, the non-linear coefficient is quantized, while the linear part is not.
From this point of view, it is expected that the four dimensional anomalous contribution can be found in the four-dimensional QH system.

\subsubsection*{Acknowledgments}

The author would like to thank Y.~Hidaka, S.~Hikami, Y.~Hirono, H.~Katsura, Y.~Minami, S.~Murakami, H.~Shimada, A.~Shitade and T.~Yoshimoto for valuable comments and discussions.
This work has been supported in part by ``Investissements d'Avenir'' program, Project ISITE- BFC (No.~ANR-15-IDEX-0003), EIPHI Graduate School (No.~ANR-17-EURE-0002) and Bourgogne-Franche-Comté Region.

\bibliographystyle{utphys}
\bibliography{conf}

\end{document}